\documentclass[11pt,english]{article}
\usepackage[T1]{fontenc}
\usepackage[latin9]{inputenc}
\usepackage{units}
\usepackage{amsthm}
\usepackage{amsmath}
\usepackage{amssymb}
\usepackage{graphicx}

\makeatletter
  \theoremstyle{plain}
  \newtheorem*{cor*}{\protect\corollaryname}
  \theoremstyle{plain}
  \newtheorem*{thm*}{\protect\theoremname}
  \theoremstyle{plain}
  \newtheorem*{lem*}{\protect\lemmaname}

\usepackage{fullpage}
\date{}

\makeatother

\usepackage{babel}
  \providecommand{\corollaryname}{Corollary}
  \providecommand{\lemmaname}{Lemma}
  \providecommand{\theoremname}{Theorem}

\begin{document}

\title{Bounding the seed length of Miller and Shi's unbounded randomness
expansion protocol}

\author{Renan Gross and Scott Aaronson}
\maketitle
\begin{abstract}
Recent randomness expansion protocols have been proposed which are
able to generate an unbounded amount of randomness from a finite amount
of truly random initial seed. One such protocol, given by Miller and
Shi, uses a pair of non-signaling untrusted quantum mechanical devices.
These play XOR games with inputs given by the user in order to generate
an output. Here we present an analysis of the required seed size,
giving explicit upper bounds for the number of initial random bits
needed to jump-start the protocol. The bits output from such a protocol
are $\varepsilon$-close to uniform even against quantum adversaries.
Our analysis yields that for a statistical distance of $\varepsilon=10^{-1}$
and $\varepsilon=10^{-6}$ from uniformity, the number of required
bits is smaller than 225,000 and 715,000, respectively; in general
it grows as $O(\log\frac{1}{\varepsilon})$.
\end{abstract}

\section{Introduction}

Building a device that generates a random string using quantum mechanics
is easy: All it needs to do is to prepare a qubit in a state in the
X basis, and then measure it in the Z basis. However, what if you
didn't build the device yourself, but instead it was given to you
by your arch-nemesis? How could you certify that the output is indeed
random, and not, for example, deterministically fixed, or somehow
correlated to the arch-nemesis? To treat these problems, protocols
have been developed recently which allow one to certify that the output
of an untrusted device was indeed random (R. Colbeck, \cite{Colbeck2006}).
The devices in these protocols don't just prepare and measure qubits
in different states, but are made of components that play XOR games
with each other. As these games require random bits as input, effort
was invested in generating more randomness than was invested as input;
this is called randomness expansion. Both polynomial (S. Pironi et.
al \cite{PAM2010}) and exponential (Vaziarni and Vidick \cite{VaziraniVidick2012})
expansion has been described, and recently even infinite expansion
(Chung, Shi, Wu and Miller \cite{ChungShiWu2014,MillerShi2014}, Coudron
and Yuen \cite{CoudronYuen2014}). The latter protocols take as input
a finite truly random string and output a nearly uniform string of
arbitrary length; further, the distance from uniformity depends only
on the number of random bits used as seed. All of \cite{ChungShiWu2014,CoudronYuen2014,MillerShi2014}
rely on a ``spot checking'' technique by Vazirani and Vidick \cite{VaziraniVidick2012}
and Coudron, Vidick and Yuen \cite{CVY2013} when generating inputs
to the XOR games; their contribution and differences are in how they
compose the inputs and outputs between devices and the analysis of
that composition. The proofs given in those papers are asymptotic
and do not give concrete bounds on the number of initial random bits
needed in order to provide this infinite expansion with the desired
soundness and security. In this paper, we give a rough estimate of
the number of bits needed in order to obtain a desired distance from
uniformity. We follow the analysis of Miller and Shi \cite{MillerShi2014},
who, in conjunction with Chung, Shi, and Wu \cite{ChungShiWu2014},
built a protocol which uses only two devices. This protocol appears
to be simpler to analyze than the one suggested by Coudron and Yuen
\cite{CoudronYuen2014} and is likely to have smaller constant overhead:
it uses less quantum devices and does not use the relatively complicated
Reichardt-Unger-Vazirani protocol \cite{RUV2012}. 

\begin{figure}
\begin{centering}
\includegraphics[scale=0.5]{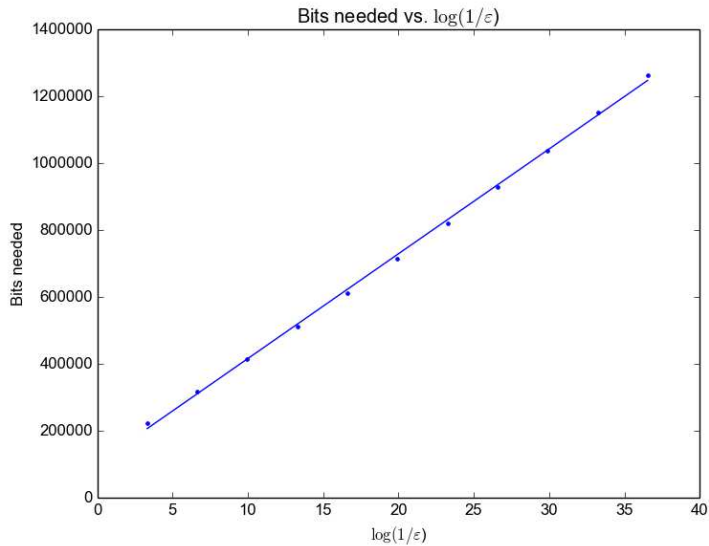}
\par\end{centering}

\caption{The seed length obtained by using the technique described in Section
\ref{sec:Multiple-iterations}. The linear dependence was obtained
by passing a fit through the needed seed length for different $\varepsilon$'s
in the range $10^{-1}$ to $10^{-11}$.\label{fig:seed_vs_eps}}
\end{figure}

The analysis yielded an upper bound for a single iteration of the
protocol - which gives exponential expansion - and a technique for
numerically approximating the number of bits needed for unbounded
iterations. For example, for an error of $10^{-6}$, the seed length
is bounded from above by 715,000; generally, from numerical calculations,
the relation between the seed length approximation and the error can
be bounded from above by a linear relation: $\text{SeedLen}\propto31328\cdot\log_{2}\frac{1}{\varepsilon}$,
as can be seen in Figure \ref{fig:seed_vs_eps}.

Randomness is needed for two purposes: generating XOR games, and extracting
randomness from high entropy strings. As we will see later, generating
the XOR games is the more demanding of the two, taking the larger
portion of the overall random bits needed: the ratio between them
is about 2:1.

\section{Spot Checking protocol}

\subsection{The gist of the protocol}

The protocol requires two identical non-interacting \emph{devices}.
A device consists of $n$ non-interacting \emph{components}; these
components are going to play a XOR game. The number $n$ depends on
the game being played; thus the CHSH game requires two components,
while the GHZ requires components. A single run consists of having
a \emph{single} device play a very large number of games. If the device
wins enough games, the protocol succeeds and an output is generated
according to its answers; by an appropriate variant of the Bell inequality
(depending on the game), this output is guaranteed to have some min-entropy.
If the device doesn't win enough games, the protocol aborts. 

The main point is that the input to the games is not uniformly random;
in fact, most of the time, the input is just zeros. Only on a small,
randomly selected number of games are the inputs chosen at random.
As the device doesn't know where the randomized inputs are going to
be, this forces it ``play honest'' and play a non-deterministic
winning strategy on almost all of the games, if it wants to pass the
protocol. Thus, a string with high min-entropy can be obtained, while
using very little random bits - most of the inputs were predetermined
zeros. 

The output string can then be fed into a randomness extractor, yielding
a nearly-uniform random string. The above procedure is then repeated
again and again, each time using the extracted string as a source
of randomness for choosing where the non-zero games are, and each
time alternating between the two devices (the alternation is an important
part of proving the quantum security, but is not needed in our analysis,
and we will not go in detail about it here).

\subsection{More formally}

The protocol is composed of \emph{iterations}. During an iterations
a single device is used, and all of its components play the XOR game.
The following arguments are fixed:
\begin{description}
\item [{\textmd{$N$:}}] The \textbf{output length}. This is a positive
integer which denotes how many times we will play the game.
\item [{$\eta:$}] The \textbf{error tolerance}. This is a real number$\in(0,\frac{1}{2})$
which denotes how large a statistical error we allow our components
to make relative to the optimal winning strategy's expectation. 
\item [{\textmd{$q$:}}] The \textbf{test probability}. This is a real
number$\in(0,1)$ which denotes the probability that a round will
be a randomized ``game round'' (see ahead).
\end{description}
The single iteration protocol, denoted $R$, is then as follows:
\begin{enumerate}
\item Repeat steps 2-4 $N$ times:
\item A bit $g\in\left\{ 0,1\right\} $ is chosen according to the distribution
$\left(1-q,q\right)$
\item If $g=1$ (``game round''), then an input string is chosen at random
from $\left\{ 0,1\right\} ^{n}$, according to the specific game chosen.
For example, for the GHZ game, the possible strings are 000, 100,
010, 001. If the devices win, record 0. Else, record 1, and mark ``Failure''.
\item If $g=0$ (``generation round''), then the input string composed
entirely of zeros $00...0$ is given to the components. Record the
bit generated by the first component.
\item If the total number of failures exceeds $(1-\mathbf{w}_{G}+\eta)qN$,
where $\mathbf{w}_{G}$ is the winning probability for the optimal
strategy, the protocol aborts. Otherwise it succeeds, and outputs
the $N$-bit sequence of outcomes it recorded.
\end{enumerate}
It can be shown that with the right choice of parameters $\eta,q$,
and $N$, the output string can be $\varepsilon$-close to $(1-\delta)N$
min-entropy for any choice of $\delta$, with $\varepsilon$ exponentially
small as a function of $N$. The amount of randomness needed to generate
this string goes roughly as $\log N$, so it is possible to generate
a string with arbitrarily more min-entropy than what we started with. 

A quantum-secure extractor is then applied to the output, yielding
a smaller but nearly uniform random string. It is possible to construct
extractors that extract a constant fraction of min-entropy, while
using an additional seed of size $O(\log^{2}\frac{N}{\varepsilon}$),
where $\varepsilon$ is the distance to uniformity%
\footnote{A quantum secure extractor is needed only if we are afraid that an
adversary might be entangled with the internal mechanism of the device,
and thus gain information about our random string. If this is not
the case - if we only wish to verify that the device generates random
bits without conditioning on a possible adversary's information -
then a constant fraction extractor can be built with a seed size of
only $O(\log\frac{N}{\varepsilon})$ \cite{DvirWigderson08}.%
}. 

Thus, running one iteration and applying an extractor yields exponentially
many more bits than we started with. By alternating between devices
and using the output of one device as the randomness seed for the
game generation and extractor of the other, an unbounded amount of
random bits can be produced.

\subsection{Layout}

We start by analyzing the seed length needed for a single iteration
of the protocol: given a target error in uniformity, how many random
bits do we need in order to get just the exponential expansion for
one device? We then look at how the error grows when we play several
iterations of the protocol. The XOR game used by the devices has been
chosen to be the GHZ game, for several reasons: it features a large
gap between the best quantum strategy (100\% win rate) and the best
classical strategy (75\% win rate); its best strategy always wins;
and Miller and Shi give a bound to its ``trust coefficient'', a
constant that appears in their analysis.

\section{Single iteration with extraction}

A single iteration requires randomness in two places: choosing the
inputs, $g_{i}$ to the XOR game for the device, and seeding the extractor.
These two are not quite independent of each other: if we play $N$
games with the device, our output will be a string of length $N$
with min-entropy linear in $N$. Both the game input randomness, and
the extractor seed length are polylogarithmic in $N$. We will start
by analyzing the XOR game, and then proceed to the extractor seed.

\subsection{Definitions}

We follow the same notation as Miller and Shi. Logarithms written
as $\log x$ are in base 2; logarithms written as $\ln x$ are natural.

For a given XOR game $G$, let $\mathbf{f}_{G}$ be smallest \emph{failing
probability} for a game; that is, the probability that the best quantum
strategy will fail to win the given game. For the GHZ game, we have
$\mathbf{f}_{G}=0$. The \emph{trust coefficient }for a game is a
number $\mathbf{v}_{G}\in\left(0,1\right]$ (described in more detail
in the main text, but no more than this is needed). For the GHZ game,
it was proven that $\mathbf{v}_{G}\geq0.14$; we will denote this
bound as $c_{v}=0.14$.

The following functions appear in the theorems and corollaries: 

\[
\Pi(x,y)\triangleq1-\left(\dfrac{1+2x}{x}\right)\log\left[\left(1-y\right)^{\frac{1}{1+2x}}+y^{\frac{1}{1+2x}}\right]
\]

\[
\pi(x)\triangleq1+2x\log x+2(1-x)\log x=1-2h(x)
\]

It can be shown that $\lim_{x\rightarrow0}\Pi(x,y)=\pi(y)$. The derivative
of the $\pi$ function is

\[
\pi'(y)=2(\log(1-y)-\log y)
\]

For $y\leq0.5$ we have:
\begin{itemize}
\item $\pi(y)$ is non-increasing and has a minimum of $-1$ at $y=0.5$.
\item $\pi'(y)$ is negative and non-decreasing, with a zero at $y=0.5$.
It tends to $-\infty$ for $y\rightarrow0$.
\end{itemize}
$H_{min}^{\varepsilon}(S)$ denotes the $\varepsilon$-smooth min-entropy
of a quantum state $S$. The state $\Gamma_{EGIO}^{s}$ denotes the
state of success of a iteration for a given adversary $E$, game $G$,
input $I$ and output $O$.

\subsection{Choosing randomness for the XOR game}

The corollary numbering in this section is according Miller and Shi's
paper \cite{MillerShi2014}. An important result in that paper is
\textbf{Corollary I.5}, which states:
\begin{cor*}[\textbf{I.5}]
 Let $\delta>0$ be a real number. Then, there exists positive reals
$K,b,q_{0}$ and $\eta$ such that the following holds. If Protocol
R is executed with parameters $N,\eta,q,$ where $q\leq q_{0}$, then
\[
H_{min}^{\varepsilon}\left(\Gamma_{EGIO}^{s}|EGI\right)\geq N\cdot\left(1-\delta\right)
\]
where $\varepsilon=K\cdot2^{-bqN}$.
\end{cor*}
This is done by proving lower bounds on the rate of entropy. To quote
Miller and Shi: ``Our approach, broadly stated, is as follows: we
show the existence of a function $T(v,h,\eta,q,k)$ which provides
a lower bound of the linear rate of entropy of the protocol. {[}...{]}
In principle, our proofs could be used to compute an explicit formula
for the function $T$, but we have not attempted to do this because
the formula might be very complicated.''

In light of these words, they have only shown the behavior of $T$
in the limit of small $q$ and $k$ parameters. In the following section,
we will:
\begin{itemize}
\item Attempt to find explicit bounds for $T$ for ``small enough'' parameters. 
\item Calculate how many random bits are needed, given fixed $q$ and $N$,
in order to obtain desired min-entropy rate and smoothness values.
\end{itemize}

\subsubsection{Bounding the $T$ function}

In the original paper, the $T$ function is given by composition of
a sequence of other functions. It can be expanded to yield:

\begin{eqnarray*}
T(v,h,\eta,q,k) & \triangleq & -\frac{1}{r_{0}qk}\cdot\underset{t\in[0,1]}{\max}[\log\Bigg(\left(1-q\right)2^{-rqk\cdot\Pi(r_{0}qk,t)}+\\
 & + & q\left\{ 1-(1-2^{-k})\left[\left(\dfrac{h}{2}\right)^{1+r_{0}qk}+v^{1+r_{0}qk}t\right]\right\} \Bigg)-\dfrac{\frac{h}{2}+\eta}{r_{0}}
\end{eqnarray*}
where 

\[
r_{0}\triangleq\min\left\{ \dfrac{-v}{\pi'(\frac{\eta}{v})},\dfrac{1}{qk}\right\} .
\]
Effectively, because $q$ and $k$ can be made arbitrarily small,
$r_{0}=\dfrac{-v}{\pi'(\frac{\eta}{v})}$. We also define:

\[
E(v,h,\eta,q,k)\triangleq\dfrac{2}{r_{0}}
\]
Which, under our assumption, simplifies to:

\[
E(v,h,\eta,q,k)=\dfrac{-2\pi'(\frac{\eta}{v})}{v}
\]
Miller and Shi use these functions to talk about the min-entropy found
in the output of a large number of games:
\begin{thm*}[\textbf{I.1}]
Suppose Protocol R is executed with parameters $N,\eta,q$. Then
for any $k\in(0,\infty)$and $\varepsilon\in(0,\sqrt{2}]$, the following
holds:

\[
H_{min}^{\varepsilon}\left(\Gamma_{EGO}^{s}|EG\right)\geq N\cdot T(\mathbf{v}_{G},2\mathbf{f}_{G},\eta,q,k)-\left(\frac{\log(\sqrt{2}/\varepsilon)}{qk}\right)E(\mathbf{v}_{G},2\mathbf{f}_{G},\eta,q,k).
\]

Also,
\[
\lim_{(q,k)\rightarrow(0,0)}T(v,h,\eta,q,k)=\pi(\frac{\eta}{v})
\]
\[
\lim_{(q,k)\rightarrow(0,0)}E(v,h,\eta,q,k)=\dfrac{-2\pi'(\frac{\eta}{v})}{v}.
\]

\end{thm*}
We will now look at the appropriate theorems from their papers and
root out the needed constants from their proofs.
\begin{cor*}[\textbf{I.2}]
 for every $\eta>0$ and $\delta>0$, there exist $b>0$, $q_{0}>0$
such that the following holds: if one iteration of the protocol is
played with parameters $N,\eta,q\leq q_{0}$, then
\[
H_{min}^{\varepsilon}\left(\Gamma_{EGO}^{s}|EG\right)\geq N\cdot\left(\pi(\frac{\eta}{\mathbf{v}_{g}})-\delta\right)
\]
and $\varepsilon=\sqrt{2}\cdot2^{-bqN}$.
\end{cor*}
The proof follows by finding $q_{0},k_{0}>0$ small enough, and $M$
large enough so that for all $q<q_{0},k<k_{0}$:

\[
T(\mathbf{v}_{G},2\mathbf{f}_{G}=0,\eta,q,k)\geq\pi(\frac{\eta}{\mathbf{v}_{g}})-\frac{\delta}{2}
\]

\[
E(\mathbf{v}_{G},2\mathbf{f}_{G}=0,\eta,q,k)\leq M.
\]
Setting $b=\dfrac{k_{0}\cdot\delta}{2M}$ then yields the correct
result. We will now find such $q_{0},k_{0}$. 

We'll start with $E$. By definition:

\[
E=\dfrac{2}{r_{0}}=\dfrac{2}{\min\left\{ \dfrac{-\mathbf{v}_{G}}{\pi'(\frac{\eta}{\mathbf{v}_{G}})},\dfrac{1}{qk}\right\} }=2\max\left\{ qk,\dfrac{-\pi'(\frac{\eta}{\mathbf{v}_{G}})}{\mathbf{v}_{G}}\right\} 
\]
So for small enough $q_{0}$ and $k_{0}$, specifically:

\[
q_{0}\cdot k_{0}\leq\dfrac{-\pi'(\frac{\eta}{\mathbf{v}_{G}})}{\mathbf{v}_{G}}\text{,}
\]
we have that

\[
E=\dfrac{-2\pi'(\frac{\eta}{\mathbf{v}_{G}})}{\mathbf{v}_{G}}
\]
We don't know $\mathbf{v}_{G}$, but $q_{0}$ and $k_{0}$ will obey
this inequality if:

\[
q_{0}\cdot k_{0}\leq-\pi'(\frac{\eta}{c_{v}})\leq\dfrac{-\pi'(\frac{\eta}{c_{v}})}{\mathbf{v}_{G}}\leq\dfrac{-\pi'(\frac{\eta}{\mathbf{v}_{G}})}{\mathbf{v}_{G}}.
\]
In this case, we have:

\[
E=\dfrac{-2\pi'(\frac{\eta}{\mathbf{v}_{G}})}{\mathbf{v}_{G}}\leq\dfrac{-2\pi'(\eta)}{c_{v}}\triangleq M
\]
And we have found our $M$. 

Satisfying the condition for $T$ requires a bit more calculations.
First we replace $\mathbf{v}_{G}$ by either 1 or $c_{v}$ as appropriate,
as in the above inequalities. We wish to make the left hand side of
the inequality $T(\mathbf{v}_{G},2\mathbf{f}_{G}=0,\eta,q,k)\geq\pi(\frac{\eta}{\mathbf{v}_{g}})-\frac{\delta}{2}$
smaller, so if we manage to solve that inequality, we will certainly
solve the original one.

We therefore look for a $q_{0},k_{0}$ pair such that for all $q<q_{0},k<k_{0}$,

\begin{eqnarray*}
\frac{-1}{r_{1}qk}\cdot\underset{t\in[0,1]}{\max}\left[\log\left(\left(1-q\right)2^{-r_{1}qk\cdot\Pi(r_{1}qk,t)}+q\left\{ 1-\left(1-2^{-k}\right)c_{v}^{1+r_{1}qk}t\right\} \right)\right]-\dfrac{\eta}{r_{1}} & \geq & \pi(\dfrac{\eta}{\mathbf{v}_{G}})-\frac{\delta}{2}
\end{eqnarray*}
where $r_{1}=\frac{-c_{v}}{\pi'(\eta)}\leq r_{0}=\frac{-\mathbf{v}_{G}}{\pi'(\frac{\eta}{\mathbf{v}_{G}})}$.

Note that $\pi(x)$ is a decreasing function in the interval $[0,\frac{1}{2}]$,
so $\pi(\dfrac{\eta}{\mathbf{v}_{G}})\leq\pi(\eta)$ as $\mathbf{v}_{G}\leq1$.
Out inequality will be satisfied if we can satisfy:

\begin{eqnarray*}
\frac{-1}{r_{1}qk}\cdot\underset{t\in[0,1]}{\max}\left[\log\left(\left(1-q\right)2^{-r_{1}qk\cdot\Pi(r_{1}qk,t)}+q\left\{ 1-\left(1-2^{-k}\right)c_{v}^{1+r_{1}qk}t\right\} \right)\right] & \geq & \pi(\eta)-\frac{\delta}{2}+\dfrac{\eta}{r_{1}}
\end{eqnarray*}

\begin{figure}
\begin{centering}
\includegraphics[scale=0.5]{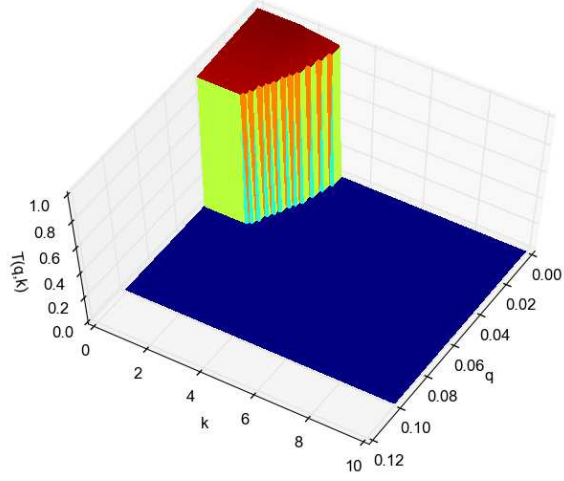}
\par\end{centering}

\caption{A 3d plot of the value of a clipped $T$ as a function of $q$ and
$k$: the value is zero if it doesn't satisfy the inequality $T(\mathbf{v}_{G},2\mathbf{f}_{G},\eta,q,k)\geq\pi(\frac{\eta}{\mathbf{v}_{g}})-\frac{\delta}{2}.$
As can be seen, only small $k$ and $q$ values are applicable. Here,
$\delta=0.025$ and $\eta=4.2\cdot10^{-5}$ \label{fig:find_q_k}}
\end{figure}

This is not easy to do analytically, but numerical calculations can
be performed. They show that for each value of $\delta$ and $\eta$,
there is only a small region around $(0,0)$ in which $q$ and $k$
can take values. An example for $\delta=0.025$, $\eta=4.2\cdot10^{-5}$
can be seen in Figure \ref{fig:find_q_k}.

It is possible to take any combination of $k_{0}$ and $q_{0}$ within
the specified range. We numerically optimize over such values to find
the pair that yields the smallest seed size.

For a given $k_{0}$ and using the $M$ previously found, we have:

\[
b=\dfrac{k_{0}\cdot\delta}{2M}=\dfrac{k_{0}\cdot\delta}{2\dfrac{-2\pi'(\eta)}{c_{v}}}=\dfrac{k_{0}\cdot\delta c_{v}}{-4\pi'(\eta)}
\]

Next, we will find the constants implied in \textbf{Corollary I.3},
which shows that there is high min-entropy even conditioned on the
input to the device.
\begin{cor*}[\textbf{I.3}]
 For every $\eta>0$ and $\delta>0$, there exist $K,>0$, $b>0$,
$q_{0}>0$ such that the following holds: if one iteration of the
protocol is played with parameters $N,\eta,q\leq q_{0}$, then
\[
H_{min}^{\varepsilon}\left(\Gamma_{EGIO}^{s}|EGI\right)\geq N\cdot\left(\pi(\frac{\eta}{\mathbf{v}_{g}})-\delta\right)
\]

\end{cor*}
As in the proof for the corollary, we choose $\delta'=\delta/2$.
The associated $q_{0}'$ and $b'$ give:

\[
\varepsilon'=\sqrt{2}\cdot2^{-b'qN}=\sqrt{2}\cdot2^{\frac{k_{0}\cdot\delta c_{v}}{4\pi'(\frac{\eta}{c_{v}})}Nq}
\]
The extra error is, assuming that $q_{0}\leq\frac{\delta}{2n}$:

\[
e^{-N\left[\delta/2n-q_{0}\right]^{2}/2}
\]
so the total error is now:

\begin{eqnarray*}
\varepsilon & = & \sqrt{2}\cdot2^{-b'qN}+e^{-N\left[\delta/2n-q_{0}\right]^{2}/2}\\
 & = & \sqrt{2}\cdot2^{-b'qN}+2^{-\frac{\log e}{2}\left[\delta/2n-q_{0}\right]^{2}N}
\end{eqnarray*}
Assuming $b'$ as before:

\[
=\sqrt{2}\cdot2^{\frac{k_{0}\cdot\delta c_{v}}{4\pi'(\eta)}qN}+2^{-\frac{\log e}{2}\left[\delta/2n-q_{0}\right]^{2}N}
\]
We want to bound this from above by a single exponent of the form
$K\cdot2^{-bqN}$. Note that

\[
2^{-\frac{\log e}{2}\left[\delta/2n-q_{0}\right]^{2}N}\leq2^{-\frac{\log e}{2}\left[\delta/2n-q_{0}\right]^{2}N\frac{q}{q_{0}}}
\]
as $q\leq q_{0}$ and the exponent would be smaller in magnitude.
Taking

\[
b=\min\left\{ \frac{k_{0}\cdot\delta c_{v}}{-4\pi'(\eta)},\frac{\log e}{2}\left[\delta/2n-q_{0}\right]^{2}\dfrac{1}{q_{0}}\right\} ,
\]
we then have:

\begin{eqnarray*}
\varepsilon & = & \sqrt{2}\cdot2^{-\frac{k_{0}\cdot\delta c_{v}}{-4\pi'(\eta)}qN}+2^{-\frac{\log_{2}e}{2}\left[\delta/2n-q_{0}\right]^{2}N}\\
 & \leq & \sqrt{2}\cdot2^{-bqN}+2^{-bqN}\\
 & = & \left(\sqrt{2}+1\right)2^{-bqN}
\end{eqnarray*}
So $K=\sqrt{2}+1$.

Next, we find the constants needed for \textbf{Corollary I.5}. We
restate it here:
\begin{cor*}[\textbf{I.5}]
 Let $\delta>0$ be a real number. Then, there exists positive reals
$K,b,q_{0}$ and $\eta$ such that the following holds. If Protocol
R is executed with parameters $N,\eta,q,G,D$, where $q\leq q_{0}$,
then
\[
H_{min}^{\varepsilon}\left(\Gamma_{EGIO}^{s}|EGI\right)\geq N\cdot\left(1-\delta\right)
\]
where $\varepsilon=K\cdot2^{-bqN}$.
\end{cor*}
In order to do so, we find an $\eta$ such that for a given $\delta$,
we have:

\[
\left|1-\pi(\frac{\eta}{\mathbf{v}_{G}})\right|\leq\frac{\delta}{2}
\]

And then when we choose $\delta'=\delta/2$ of the original, we are
guaranteed to be within the range $[1-\delta,1]$, as needed for the
corollary. Remembering that $\pi(x)$ is a decreasing function of
$x$, we can set $\mathbf{v}_{G}=c_{v}$, and the inequality will
still hold. Lets mark $x=\frac{\eta}{c_{v}}$ and look at the behavior
of $\pi(x)$. We want the following inequality to hold:

\begin{eqnarray*}
1-\pi(x) & \leq & \delta/2
\end{eqnarray*}

Opening up the $\pi$ function, this reduces to:

\[
-x\log x-(1-x)\log(1-x)\leq\delta/4
\]

This can easily be found numerically, yielding a number $x_{0}\in(0,0.5)$.
We then have:

\[
\eta_{max}=x_{0}\cdot c_{v}
\]

With this we generate an $\eta\in(0,\eta_{max})$ parameter and the
constants $q_{0}$, $k_{0}$, $K$, and $b$ (with a $\delta$ value
one fourth of the one we used for calculating $\eta$, as we had to
halve it twice in our proofs). For these parameters, playing one iteration
of the expansion protocol will produce a string with min-entropy:

\[
H_{min}^{\varepsilon}\left(\Gamma_{EGIO}^{s}|EGI\right)\geq N\cdot\left(1-\delta\right)
\]
where the smoothness is bounded by: 

\[
\varepsilon=K\cdot2^{-bqN}
\]
Conversely, for a given $\varepsilon$ of required smoothness, we
have the following constraint on $q$ and $N$:

\[
q\cdot N=-\frac{\log\frac{\varepsilon}{K}}{b}=\frac{\log\frac{K}{\varepsilon}}{b}
\]
This will be used when deciding on $N$ and $q$ values for a desired
error level.

\subsubsection{Random bits for protocol R\label{sub:Random-bits-for}}

Having established a relation between $q$ and $N$, we can proceed
to calculate the number of random bits needed in order to execute
protocol $R$ with $N$ games.

Randomness comes into play in two places in protocol $R$: when deciding
on which games we use random inputs instead of dummy zeros, and choosing
the actual inputs when this happens. Since we are playing the GHZ
game, the latter requires 2 bits for each time we play a real game. 

By definition of the bits $g_{i}$, generating them requires no more
random bits than their Shannon entropy. Combining this with the previous
statement, we need 
\[
2N\left(-q\log q-(1-q)\log(1-q)\right)
\]
initial random bits in order to play $N$ games.

\subsection{Seed length for the extractor}

\subsubsection{Quantum secure extractor\label{sub:Quantum-secure-extractor}}

Part of the initial randomness needed for one execution of the protocol
is the random seed given to the extractor, which we apply on our $N$
bit output that came from playing $N$ games. Based on the paper ``Trevisan's
extractor in the presence of quantum side information'' \cite{De2012},
we will construct, from bottom up, a suitable extractor. For the purpose
of this analysis, we assume that out extractor will operate on $N$
bits which have at least a constant fraction of $\varepsilon$-smooth
min-entropy.

Trevisan's extractor and its quantum security relies on single bit
extractors; the ones in \cite{De2012} use list-decodable codes. All
theorems and lemmas in this section are numbered according to \cite{De2012}.
\begin{lem*}[\textbf{C.2}]
 For every $n\in\mathbb{N}$ and $\delta>0$, there is code $C_{n,\delta}:\left\{ 0,1\right\} ^{n}\rightarrow\left\{ 0,1\right\} ^{\bar{n}}$
that is $(\delta,\frac{1}{\delta^{2}})$-list-decodable. Further,
$\bar{n}=\text{poly}\left(n,\frac{1}{\delta}\right)$.
\end{lem*}
Guruswami et al. \cite{Guruswami00} give a construction with $\bar{n}=O\left(\frac{n}{\delta^{4}}\right)$;
after extracting the constants we have

\[
\bar{n}\leq32\frac{n}{\delta^{4}}.
\]

Knowing how to construct list-decodable codes, we can use them as
extractors:
\begin{thm*}[\textbf{C.3}]
 Let $C:\left\{ 0,1\right\} ^{n}\rightarrow\left\{ 0,1\right\} ^{\bar{n}}$
be an $(\varepsilon,L)$ list-decodable code. Then 
\begin{eqnarray*}
C':\left\{ 0,1\right\} ^{n}\times\left[\bar{n}\right] & \rightarrow & \left\{ 0,1\right\} \\
\left(x,y\right) & \mapsto & C(x)_{y}
\end{eqnarray*}
is a $\left(\log L+\log\frac{1}{2\varepsilon},2\varepsilon\right)$-strong
extractor.
\end{thm*}
Notice that as an extractor, the seed given to $C'$ has $\bar{n}$
different inputs, and therefore requires only $t=\log\bar{n}$ bits
of randomness.

Combining the two, for any $\varepsilon>0$ we can build an extractor
by putting in $\delta=\frac{\varepsilon}{2}$ into \textbf{Lemma C.2},
and using that list-decodable code in \textbf{Theorem C.3}. This will
give a $(3\log\frac{1}{\varepsilon},\varepsilon)$ extractor (actually,
we have a $(\log\frac{1}{4\varepsilon^{2}}+\log\frac{1}{4\varepsilon},\varepsilon)$-extractor,
but $\log\frac{1}{4\varepsilon^{2}}+\log\frac{1}{4\varepsilon}=2\log\dfrac{1}{2\varepsilon}+\log\frac{1}{4\varepsilon}=3\log\frac{1}{\varepsilon}-6$,
so we certainly have a $(3\log\frac{1}{\varepsilon},\varepsilon)$-extractor
as well).

Our extractor requires $t=\log\left(32\frac{16n}{\varepsilon^{4}}\right)=\log\left(512\frac{n}{\varepsilon^{4}}\right)$
bits of randomness. 

Next we will compose 1-bit extractors to create general ones:
\begin{thm*}[\textbf{4.6}]
 Let $C:\left\{ 0,1\right\} ^{n}\times\left\{ 0,1\right\} ^{t}\rightarrow\left\{ 0,1\right\} $
be a $(k,\varepsilon)$-strong 1-bit extractor with uniform seed,
and $S_{1},...,S_{m}\subset[d]$ a weak $(t,r)$-design. Then a Trevisan
style extractor composition, $\text{Ext}_{C}:\left\{ 0,1\right\} ^{n}\times\left\{ 0,1\right\} ^{d}\rightarrow\left\{ 0,1\right\} ^{m}$
gives a $(k+rm+\log\frac{1}{\varepsilon},3m\sqrt{\varepsilon})$-quantum-proof-strong
$(n,d,m)$ extractor. 
\end{thm*}
We are going to use this theorem in order to get a constant rate extractor.
Assume that you want an extractor whose output is $\varepsilon$-close
to uniform. Then the 1-bit extractor needs to be $\varepsilon'=\frac{\varepsilon^{2}}{9m^{2}}$
close to uniform. Set $k=m$ and $r=2\frac{k}{m}=2$. Then the resultant
extractor produces output that is $\varepsilon$ away from uniform
and works with any entropy larger than

\begin{eqnarray*}
k+rm+\log\frac{1}{\varepsilon'} & = & k+2m+\log\frac{9m^{2}}{\varepsilon^{2}}\\
 & = & k+2k+2\log\frac{3m}{\varepsilon}\\
 & = & 3k+2\log\frac{3k}{\varepsilon}\\
 & = & 3k+2\log\frac{1}{\varepsilon}+2\log3k\\
 & \leq & 4k+2\log\frac{1}{\varepsilon}
\end{eqnarray*}

Ignoring the $\log\frac{1}{\varepsilon}$ in the min-entropy (it will
be small comparable to $k$), we get that a source with $4k$ min-entropy
can give us $m=k$ random bits. Since our $N$ bits have $(1-\delta)N$
min-entropy, we have $k=\frac{1}{4}(1-\delta)N$, and also $m=\frac{1}{4}(1-\delta)N$.
Putting this into $t$, we get:

\begin{eqnarray*}
t & = & \log(512\frac{N}{\varepsilon^{4}})\\
 & = & \log(512\frac{N}{\frac{\varepsilon^{8}}{9^{4}m^{8}}})\\
 & = & \log(512\cdot9^{4}\frac{Nm^{8}}{\varepsilon^{8}})\\
 & = & \log(512\cdot9^{4}\frac{N\left(\frac{1}{4}(1-\delta)N\right)^{8}}{\varepsilon^{8}})\\
 & = & \log(512\left(\dfrac{3}{4}(1-\delta)\right)^{8}\cdot\frac{N^{9}}{\varepsilon^{8}})\\
 & = & \log512+8\log\dfrac{3}{4}(1-\delta)+\log N+8\log\frac{N}{\varepsilon}
\end{eqnarray*}
What is $d$, and how do we get a design? According to \textbf{Lemma
5.5} in \cite{De2012}, we can build the desired design with 

\[
d=t\left\lceil \frac{t}{\ln r}\right\rceil 
\]

We chose $r=2$ so this can be effectively written as

\[
d\leq\frac{t(t+1)}{\ln2}
\]

\subsection{Total randomness for one iteration}

Choose a desired distance from uniformity $\varepsilon>0$, and the
two parameters $\delta\in\left(0,1\right)$ and $\eta\leq\eta_{max}$.
Find the constants $q_{0}$, $k_{0}$, $b$. From the relation $q\cdot N=-\frac{\log\frac{\varepsilon}{K}}{b}=\frac{\log\frac{\sqrt{2}+1}{\varepsilon}}{b}$
and the fact that $q\leq q_{0}$, we have a lower bound on $N$; pick
any $N$ greater than that, and calculate the corresponding $q$.
The total bits of randomness required is then given by the combined
result of section \ref{sub:Random-bits-for} and section \ref{sub:Quantum-secure-extractor}:

\[
\text{SeedLen}\leq2N\left(-q\log q-(1-q)\log(1-q)\right)+\frac{t(t+1)}{\ln2}
\]

This will generate $\frac{1}{4}(1-\delta)N$ bits which are $2\varepsilon$
close to uniform - one $\varepsilon$ is due to the min-entropy smoothness,
the other is due to the expander. 

This process is only fruitful if the number of generated bits is larger
than the number of input bits. This may not be the case for any choice
of $\varepsilon$, $q$, and $N$; however, we can always attain this
property by increasing $N$: For a fixed $\varepsilon$, the extractor
term grows as $\log^{2}N$, while the game generation term grows as
$-Nq\log q\propto-\log\frac{1}{N}=\log N$.

Choosing $\delta$ and $\eta$ is not trivial. A small $\delta$ value
means that the resultant string has higher min-entropy and thus more
bits are extracted; however, it also means tighter constraints for
$q_{0}$ and $k_{0}$. The parameter $\eta$ affects $b$, $q_{0}$
and $k_{0}$ in a non-linear way. Hence, we optimized these parameters
numerically for each choice of fixed $\varepsilon$.

\section{Multiple iterations and results\label{sec:Multiple-iterations}}

The errors for multiple iterations are additive: Using randomness
that is $\varepsilon$ away from uniformity for an algorithm that
expects uniform randomness will add an $\varepsilon$ to the output
error. In order to keep the error constrained, we must therefore decrease
it exponentially (or more) after each time the protocol is played. 

The simplest we can do is to cut $\varepsilon$'s value in half after
each iteration. This will give no more than $4\varepsilon$ error,
and is certainly achievable - the number of output bits grows exponentially,
while the increase in the number of bits caused by halving $\varepsilon$
is polynomial. This strategy actually overshoots, as the exponential
expansion means that there will be many bits left over after each
iteration which are not used for the next iteration. Here is an estimation
scheme based on the above:

\textbf{Scheme:} The largest seed requirement is imposed by the first
iteration. We can minimize the number of excess bits produced in the
first iteration, as follows: find a combination of $q$ and $N$ such
that the generated number of bits is \emph{just} the amount required
for the next iteration (which has $\varepsilon$ half the original,
so requires more bits). Of course, this number too requires calculation;
a simple estimation is achieved assuming that the number of bits generated
in the second iteration is in the same proportion to its seed as the
number of bits generated in the first iteration is to the initial
seed. This scheme means there is no loss of seed in the first iteration,
while there may be loss in the next ones; however, it is easy to implement.

\begin{figure}
\begin{centering}
\includegraphics[scale=0.5]{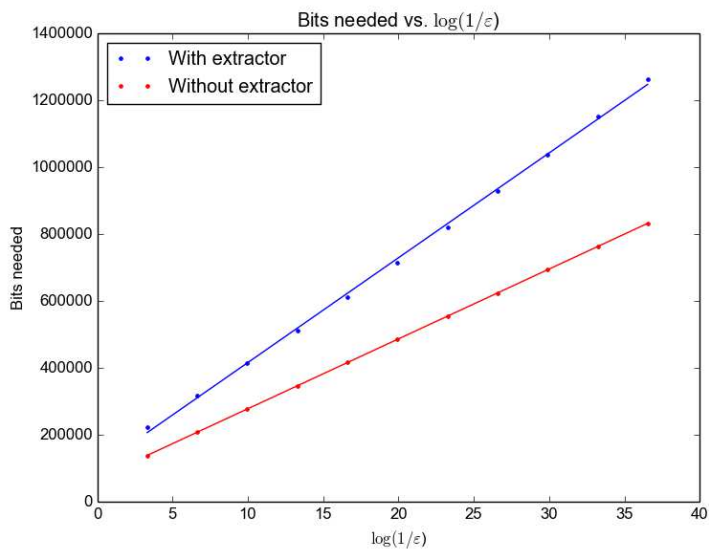}
\par\end{centering}

\caption{Amount of randomness needed vs. $\log\frac{1}{\varepsilon}.$ In blue
is the regular scheme; in red is the same scheme but assuming that
our extractor is deterministic and requires no random bits.\label{fig:with_without}}
\end{figure}

Using the first method with an initial error of $0.25\cdot10^{-6}$
(to yield a total distance of no more than $10^{-6}$ away from uniformity)
gives an initial seed of less than 715,000 bits. In general, plotting
the required number for several different $\varepsilon$ values shows
a linear relation between the seed length and $\log\frac{1}{\varepsilon}$.
This is shown in blue in Figure \ref{fig:with_without}, with a slope
of $31328$.

A presumably better technique would be to use \emph{all }the generated
bits as seed in each consecutive iteration. We set an $\varepsilon$
for the first iteration and a target $N$. Then, for each iteration,
we optimize the expression
\[
\text{SeedLen}\leq2N\left(-q\log q-(1-q)\log(1-q)\right)+\frac{t(t+1)}{\ln2}
\]
so as to get the smallest $\varepsilon$ possible. 

The analysis of putting a bound on the resultant $\sum\varepsilon_{i}$
has not been performed. Further, in order to achieve global optimization
one still has to choose an initial $N$ for the first iteration. However,
a lower bound for this value can be obtained (for this particular
approximation method) by looking at just one iteration: for a given
$\varepsilon$, how many initial bits must we use \emph{just} to get
back what we invested? As any further iterations just increase the
error, and as getting a longer output inherently requires more initial
random bits, this gives a bound from below. So running the $R$ protocol
once with $\varepsilon$ four times the desired value gives us a lower
bound. Multiplying $\varepsilon$'s value by four is equivalent to
decreasing $\log\frac{1}{\varepsilon}$ by 2. So with the current
slope obtained, this can give an improvement of no more than 64,000
bits. 

It is interesting to ask which one of the two imposes stronger requirements:
generating XOR games, or seeding the extractor. For one iteration,
the extractor seed requires $O(\log^{2}(\frac{N}{\varepsilon}))$
random bits, while the XOR game generation requires $O(Nq\log q)$.
However, $N$, $q$ and $\varepsilon$ are connected, and cannot be
changed independently. Figure \ref{fig:with_without} shows the number
of bits required as a function of $\log\frac{1}{\varepsilon}$ (applying
the same method), assuming that the extractor operates \emph{for free}
- it costs us no bits at all to extract. Of course, there are no deterministic
extractors, but this gives a bound on the XOR games. It appears that
the XOR game generation requires about $\nicefrac{2}{3}$ of the randomness
- the ratio between the two slopes is $\frac{\text{with extractor}}{\text{without}}=\frac{31875}{21380}=1.49$. 

\section{Future work}

While we have given a rough upper bound for the amount of randomness
needed to ``jumpstart'' infinite expansion, questions and research
directions still remain.
\begin{enumerate}
\item What is the optimal XOR game for such a protocol? 
\item Is the Coudron-Yuen analysis more or less efficient than the one by
Miller and Shi? Can this type of protocol be improved upon? What is
the theoretical lower bound for any infinite protocol?
\item How does the number of bits improve if we do not ask for quantum security
(security against entanglement with the inner working of the devices),
but just want to certify randomness? Are there $O(\log\frac{N}{\varepsilon})$
constant rate strong extractors which can be shown to be quantum secure?\end{enumerate}
\begin{description}
\item [{Acknowledgements}] Renan thanks Scott Aaronson for overview and
supervision. We thank Matthew Coudron for clearing things up.
\end{description}
\bibliographystyle{plain}
\addcontentsline{toc}{section}{\refname}\bibliography{gross_aaronson_bounding_seed_length}

\begin{thebibliography}{10}

\bibitem{ChungShiWu2014}
Kai-Min Chung, Yaoyun Shi, and Xiaodi Wu.
\newblock Physical randomness extractors: Generating random numbers with
  minimal assumptions.
\newblock {\em arXiv:1402.4797}, 2014.

\bibitem{Colbeck2006}
Roger Colbeck.
\newblock {\em Quantum and relativistic protocols for secure multi-party
  computation}.
\newblock PhD thesis, University of Cambridge, 2006.

\bibitem{CVY2013}
Matthew Coudron, Thomas Vidick, and Henry Yuen.
\newblock Robust randomness amplifiers: Upper and lower bounds.
\newblock {\em arXiv:1305.6626}, 2013.

\bibitem{CoudronYuen2014}
Matthew Coudron and Henry Yuen.
\newblock Infinite randomness expansion and amplification with a constant
  number of devices.
\newblock {\em Proceedings of the forty-sixth annual ACM symposium on Theory of
  computing}, pages 427--436, 2014.

\bibitem{De2012}
A.~De, C.~Portmann, T.~Vidick, and R.~Renner.
\newblock Trevisan's extractor in the presence of quantum side information.
\newblock {\em SIAM Journal on Computing}, 41(4):915--940, 2012.

\bibitem{DvirWigderson08}
Z.~Dvir and A.~Wigderson.
\newblock \text{Kakeya Sets, New Mergers, and Old Extractors}.
\newblock {\em SIAM J. on Computing}, 40(3):778--792, 2011.
\newblock (Extended abstract appeared in FOCS 2008).

\bibitem{Guruswami00}
Venkatesan Guruswami, Johan~H\aa stad, Madhu Sudan, and David Zuckerman.
\newblock Combinatorial bounds for list decoding.
\newblock {\em IEEE Transactions on Information Theory}, 48:2002, 2000.

\bibitem{MillerShi2014}
Carl~A. Miller and Yaoyun Shi.
\newblock Robust protocols for securely expanding randomness and distributing
  keys using untrusted quantum devices.
\newblock {\em arXiv:1402.0489}, 2014.

\bibitem{PAM2010}
S.~Pironio, A.~Acín, S.~Massar, A.~Boyer de~la Giroday, D.~N. Matsukevich,
  P.~Maunz, S.~Olmschenk, D.~Hayes, L.~Luo, T.~A. Manning, and C.~Monroe.
\newblock Random numbers certified by {B}ell's theorem.
\newblock {\em Nature}, 2010.

\bibitem{RUV2012}
Ben~W. Reichardt, Falk Unger, and Umesh Vazirani.
\newblock A classical leash for a quantum system: Command of quantum systems
  via rigidity of chsh games.
\newblock {\em arXiv:1209.0448}, 2012.

\bibitem{VaziraniVidick2012}
Umesh Vazirani and Thomas Vidick.
\newblock Certifiable quantum dice.
\newblock {\em Proceedings of the forty-fourth annual ACM symposium on Theory
  of computing}, pages 61--76, 2012.

\end{thebibliography}

\end{document}